\documentclass[preprint]{aastex63}

\newcommand{\gq}[1]{{#1}}
\newcommand{\gqq}[1]{{#1}}

\submitjournal{ApJL}

\shorttitle{Energy and waiting time distributions of FRB 121102 detected by FAST}
\shortauthors{Zhang et al.}

\begin{document}

\title{Energy and waiting time distributions of FRB 121102 observed by FAST}

\author[0000-0001-6545-4802]{G. Q. Zhang}
\affiliation{School of Astronomy and Space Science, Nanjing University, Nanjing 210093, China}

\author[0000-0002-3386-7159]{P. Wang}
\affiliation{CAS Key Laboratory of FAST, NAOC, Chinese Academy of Sciences, Beijing 100101, China}

\author[0000-0001-6021-5933]{Q. Wu}
\affiliation{School of Astronomy and Space Science, Nanjing University, Nanjing 210093, China}

\author[0000-0003-4157-7714]{F. Y. Wang} \affiliation{School of Astronomy and Space Science, Nanjing
    University, Nanjing 210093, China} \affiliation{Key Laboratory of
    Modern Astronomy and Astrophysics (Nanjing University), Ministry of
    Education, Nanjing 210093, China}

\author[0000-0003-3010-7661]{D. Li}
\affiliation{CAS Key Laboratory of FAST, NAOC, Chinese Academy of Sciences, Beijing 100101, China}
\affiliation{University of Chinese Academy of Sciences, Beijing 100049, People’s Republic of China}
\affiliation{NAOC-UKZN Computational Astrophysics Centre, University of KwaZulu-Natal, Durban 4000, South Africa}

\author[0000-0002-7835-8585]{Z. G. Dai}
\affiliation{Department of Astronomy, School of Physical Sciences, University of Science and Technology of China, Hefei 230026, China}
\affiliation{School of Astronomy and Space Science, Nanjing
	University, Nanjing 210093, China}
	
\author[0000-0002-9725-2524]{B. Zhang}
\affiliation{Department of Physics and Astronomy, University of Nevada, Las Vegas, Las Vegas, NV 89154, USA}

\correspondingauthor{F. Y. Wang}
\email{fayinwang@nju.edu.cn}

\begin{abstract}
	The energy and waiting time distributions are important properties \gq{for understanding}
	the physical mechanism of repeating fast radio bursts (FRBs). Recently, \gq{the} Five-hundred-meter Aperture Spherical radio Telescope (FAST, \citealt{nan11,li18fast}) detected the largest \gq{burst} sample of FRB 121102, containing 1652 \gqq{bursts} \cite{Li2021}. \gq{We use this sample to investigate the energy and waiting time distributions.} The energy \gq{count}
	distribution \gq{$dN/dE$} at the high-energy range ($>10^{38}$ erg) can be fitted with a single power-law function with an index of $-1.86 ^{+0.02}_{-0.02}$, \gq{while}
	the distribution at the low-energy range deviates from the power-law function. \gq{An interesting result of \citet{Li2021} is that there is an apparent temporal gap between early bursts (occurring before MJD 58740) and late bursts (occurring after MJD 58740).}
	\gq{We find} the energy distributions of high-energy bursts
	at different epochs are inconsistent. \gq{The power-law index is $-1.70^{+0.03}_{-0.03}$ for early bursts and $-2.60^{+0.15}_{-0.14}$ for late bursts.}
	For bursts observed in a single day, a linear repetition pattern is found. 
	We use the Weibull function to fit the distribution \gq{of waiting time of consecutive bursts.}
	The shape parameter $k = 0.72^{+0.01}_{-0.01}$ and the
	event rate $r = 736.43^{+26.55}_{-28.90}$ day$ ^{-1} $ are derived. If the waiting times with $\delta_t < 28$ s are excluded, the burst behavior can be described by a Poisson
	process. The best-fitting values of $k$ are slightly different for low-energy \gq{($E < 1.58\times10^{38}$ erg)} and high-energy \gq{($E > 1.58\times10^{38}$ erg)} bursts. 
\end{abstract}


\section{Introduction} \label{sec:intro}
Fast radio bursts (FRBs) are bright bursts at radio frequency with milliseconds-duration \citep[for reviews, see][]{Cordes2019ARA&A..57..417C,Petroff2019A&ARv..27....4P,Zhang2020,Xiao2021arXiv210104907X}.
According to their burst \gq{activity}, FRBs can be divided into two classes: repeating FRBs and apparently non-repeating FRBs.
There are more than twenty repeating FRBs detected so far, \gq{such as FRB 121102} \citep{Spitler2016Natur.531..202S} and FRB 180916 \citep{CHIME/FRBCollaboration2019ApJ...885L..24C}.
Whether all FRBs are repeating sources is still under debate \citep{Palaniswamy2018,Caleb2019,Ravi2019b,Ai2021}.
The repeating \gq{activity} excludes some \gq{cataclysmic} models. Recently, FRB 200428 was found to be associated
with a Galactic magnetar SGR 1935+2154 \citep{CHIME/FRBCollaboration2020,Bochenek2020Natur.587...59B}, which supports that at least a part of FRBs is produced by magnetars.
Some models have been proposed to interpret how magnetars may produce FRBs,
\gq{such as magnetars in supernova remnants \citep{Murase2016} or invoking shock interactions \citep{Beloborodov2017,Metzger2019}, magnetars with low magnetospheric twist \citep{Wadiasingh2019}, and crust fracturing magnetars \citep{Yang2021b}}.

FRB 121102 is the first detected repeating FRB, which has been observed from 400 MHz to 8 GHz \citep{Spitler2014ApJ...790..101S,Spitler2016Natur.531..202S,Chatterjee2017Natur.541...58C,Michilli2018Natur.553..182M,Gajjar2018ApJ...863....2G}.
In the past few years, about 300 bursts have been observed from this source. The high burst rate enables it to be investigated extensively. It was localized in a star-forming region in a dwarf galaxy in association with a persistent radio source \citep{Chatterjee2017Natur.541...58C,Marcote2017,Tendulkar2017}. The repeating bursts from FRB 121102 has \gqq{an extreme high} rotation measure (RM)  \gq{of about $10^5$ rad m$^{-2}$ \citep{Michilli2018Natur.553..182M}, which is variable and decreasing \citep{Michilli2018Natur.553..182M,Hilmarsson21}.}

The energy distribution of repeating FRBs provides a clue to understanding the physical mechanism to produce FRBs. \citet{Wang2017} first found that the differential energy distribution of FRB 121102 is $dN/dE \propto E^{-1.8\pm0.15}$. \citet{Law2017} estimated that the energy distribution of FRB 121102
satisfies a \gqq{single} power-law distribution with $dN/dE \propto E^{-1.7}$. However, \citet{Gourdji2019} derived a power-law
index of about $-2.8 \pm 0.3$. They argued that the differences between these results are caused by the incompleteness of the low energy.
\citet{Wang2019ApJ...882..108W} used the cut-off power-law function to fit bursts observed by multiple observations and found that the power-law indices of different observations are close to $-1.7$.
The Apertif data suggested a slope of $-2.7\pm0.6$ above the completeness threshold  $\sim 10^{39}$ erg \citep{Oostrum2020AA...635A..61O}.
An index $-2.1\pm0.1$ was found by \citet{Cruces2021MNRAS.500..448C} using the bursts detected by Effelsberg telescope. Therefore, the energy distribution of FRB 121102 has been controversial. 
To have a better understanding of the energy distribution, more bursts spanning a large energy range are required.

Although many bursts of FRB 121102 have been observed, the burst \gq{activity} is still confusing. A possible long-term period has been discovered for FRB 121102.
\cite{Rajwade2020MNRAS.495.3551R} derived a period of about 156.9 days. This result
was examined by \citet{Cruces2021MNRAS.500..448C}. They found a period of about 161 days, which is consistent with the result of \citet{Rajwade2020MNRAS.495.3551R}.
The similar periodic behavior (with a period of $\sim$ 16 days) was discovered earlier for FRB 180916 \citep{Chime/FrbCollaboration2020Natur.582..351C}. Many models have been proposed to explain this periodic
behavior, including binary star models \citep{Dai2016ApJ...829...27D, Ioka2020ApJ...893L..26I,Dai2020ApJ...895L...1D,Lyutikov2020ApJ...893L..39L,Deng2021,Kuerban2021,Wada2021,LiQ2021}, precession models \citep{Zanazzi2020ApJ...892L..15Z,Levin2020ApJ...895L..30L,Yang2020},
and ultra-long period magnetar models \citep{Beniamini2020MNRAS.496.3390B}.  

No shorter period has been discovered for FRB 121102 \citep{Zhang2018ApJ...866..149Z,Li2021}.
\citet{Wang2017} first found that the repetition pattern of FRB 121102 cannot be described by a Poisson process. A similar conclusion was drawn by \citet{Oppermann2018MNRAS.475.5109O}.
They also suggested that a Weibull function is suitable to describe the distribution of the waiting times \gq{of two consecutive bursts}. The shape
parameter $k$ is $0.34^{+0.06}_{-0.05}$, which means that the bursts of FRB 121102 tend to cluster in time.
\citet{Oostrum2020AA...635A..61O} also used the Weibull function to fit the burst behavior and obtained $k = 0.49 \pm 0.05$.
A higher shape parameter $k = 0.62^{+0.10}_{-0.09}$ was derived by \citet{Cruces2021MNRAS.500..448C}. \gq{The values of $k$ are slightly different for multiple observations, which may be caused by different observing times, telescope sensitivities, or unknown periodic activities. A large sample with a long observing time and high sensitivity is required to clarify this issue.}

Recently, the Five-hundred-meter Aperture Spherical radio Telescope (FAST) detected 1652 bursts from FRB 121102 \citep{Li2021}, 
which is the largest burst sample for this source. The high sensitivity of FAST \citep{li18fast} enables it to detect many low-energy bursts, which are difficult
to detect with other telescopes. This large sample is helpful to reveal the properties of FRB 121102. \citet{Li2021} reported bimodal energy distribution. More interestingly,
there are three obvious peaks in the kernel density estimation (KDE) of energy and burst time (as shown in the panel (c) of their Figure 1). The high energy peak is only visible in early observations, while the low energy peaks exist in both early and late observations \citep{Li2021}. 

In this paper, we use FAST burst sample to investigate the energy distribution and burst \gq{activity} of FRB 121102. This paper is organized as follows.
In Section 2, we introduce the data and derive the energy distribution. The Weibull function is used to fit the waiting time distribution.
The discussion is presented in Section 3 and conclusions are drawn in Section 4.

\section{Data and Results} \label{sec:method}
\subsection{Data} \label{subsec:data}
FRB 121102 has been observed by different telescopes \citep{Spitler2014ApJ...790..101S,Spitler2016Natur.531..202S,Scholz2016ApJ...833..177S,
    Scholz2017ApJ...846...80S,Chatterjee2017Natur.541...58C,Law2017,Hardy2017MNRAS.472.2800H,Zhang2018ApJ...866..149Z,Spitler2018ApJ...863..150S,
    Gajjar2018ApJ...863....2G,Michilli2018Natur.553..182M,Gourdji2019,Hessels2019ApJ...876L..23H,Oostrum2020AA...635A..61O,Caleb2020MNRAS.496.4565C,Cruces2021MNRAS.500..448C}.
Recently, FAST observed this source at 1.25 GHz \citep{Li2021}. 
The observations were carried out from August 2019 (MJD 58724) to
October 2019 (MJD 58776), a duration lasting \gqq{nearly half of the 157d period suggested by \cite{Rajwade2020MNRAS.495.3551R} and covering almost the entire active phase (MJD 58717 to 58813) predicted by \citet{Cruces2021MNRAS.500..448C}.}
During the 59.5 observing hours, FAST detected 
1652 bursts. The mean burst rate was about 27.8 hr$^{-1}$, and the peak burst rate was 122 hr$^{-1}$. These bursts help to understand the energy distribution in both the low-energy and high-energy ranges. The waiting time distribution can be studied in detail with this sample. 
We will use this sample to derive a more precise waiting time distribution.

\subsection{Energy Distribution}\label{subsec:energy}
\begin{figure}
    \centering
    \includegraphics[width = 0.8\linewidth]{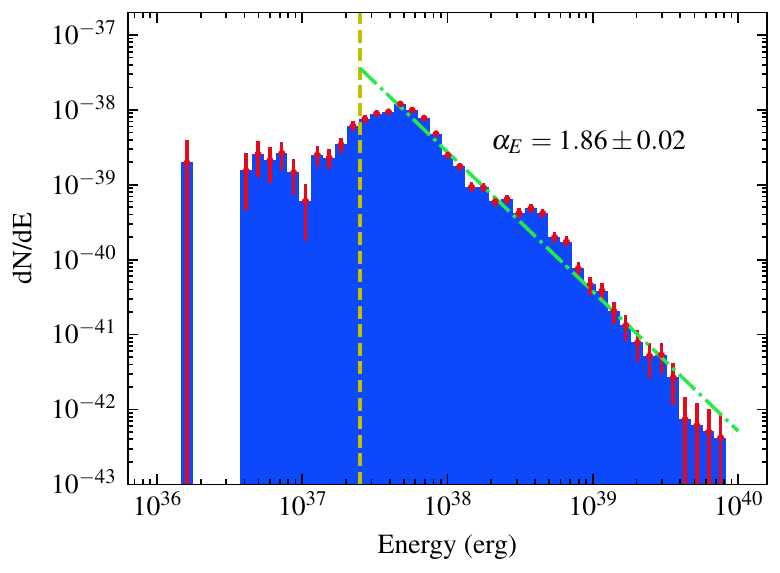}
    \caption{\gq{The differential energy distribution $dN/dE$ for the bursts of FRB 121102 observed by \citet{Li2021}.} The blue histogram is the 
    \gq{differential} energy distribution and the 
    red points and red vertical lines are the values and $1\sigma$ uncertainties. \gq{The vertical yellow dashed line is the 90\% \gq{completeness} threshold of FAST telescope \citep{Li2021}.} We use \gq{the simple power-law (equation \ref{dN})} to fit 
    the energy distribution in the high-energy range ($E>10^{38}$ erg) and show the best-fitting result as dot-dashed green line. The power-law 
    index is $-1.86\pm0.02$. 
    The bursts in the low energies deviate from power-law form.}
    \label{fig:dNdE}
\end{figure}

The energy distribution of FRB 121102 has been investigated by many previous works \citep{Wang2017,Law2017,Gourdji2019,Wang2019ApJ...882..108W,Cheng2020,Oostrum2020AA...635A..61O,Cruces2021MNRAS.500..448C,Li2021}. 
We use the bursts of FRB 121102 observed by FAST to investigate the differential energy \gq{count} distribution 
\begin{equation}\label{dN}
dN/dE\propto E^{-\alpha_E}.
\end{equation} 
The energies of these bursts are listed in \citet{Li2021}. 
\gq{They use central frequency rather than bandwidth to calculate the energies of bursts \citep[This is to alleviate the bandwidth limitation of telescopes, see][]{ZhangB2018}.} The luminosity distance of FRB 121102 is taken as 949 Mpc \citep{PlanckCollaboration2016A&A...594A..13P,Tendulkar2017}.
We show the \gqq{differential energy count} distribution as a blue histogram in Figure \ref{fig:dNdE}. 
The distribution \gqq{in the low energy range} deviates from the power-law function. 
Therefore, to compare with previous works, we only consider the high-energy bursts ($E>10^{38}$ erg), \gqq{which contains roughly 600 bursts.}
The distribution can be approximated as a power-law function. We use equation (\ref{dN}) to fit the
energy distribution in the high-energy end. The fitting result is shown as the green dot-dashed line in Figure \ref{fig:dNdE}. \gq{We also show the 90\% threshold of FAST \citep{Li2021} as vertical yellow dashed line in this figure.} The power-law index is $\alpha_E = 1.86\pm0.02$ (1$\sigma$ uncertainty), which is consistent with the index derived by \citet{Wang2017}. \citet{Law2017} obtained a power-law index of $\alpha_E \simeq 1.7$, which is also consistent with our result.
\cite{Wang2019ApJ...882..108W} used a cut-off power-law model to fit the FRB 121102 data from multiple observations and derived a universal energy distribution with $\alpha_E = 1.6-1.8$ for multiple 
observations at varied frequencies. Our result also closes to this range. The energy distribution of high-energy bursts was also investigated by \citet{Li2021}. Because the observation of FAST for 
FRB 121102 consists of many single-day observations, \citet{Li2021} used various observational times as the weight to derive the differential burst rate $dR/dE$ at different energies and found a power-law index of about $-1.85\pm0.3$. Our result is the distribution of burst counts in each energy bin $dN/dE$, not the burst rate distribution $dR/dE$.

Interestingly, the energies of non-repeating FRBs also show a power-law distribution with $\alpha_E = 1.6-1.8$ \citep{Cao2018,Lu2019,Zhangg2019,ZhangR2021,Luo18,Luo20}. Some sources also show a power-law distribution of $dN/dE$, such as X-ray bursts of magnetars \citep[\gq{e.g. $\alpha_E = 1.68 \pm 0.01$ for SGR 1806-20,}][]{Gogus2000,Prieskorn2012,Cheng2020,Yang2021}, X-ray flares of gamma-ray bursts \citep[$\alpha_E = 1.06\pm0.15$,][]{Wang2013}, M87 \citep[$\alpha_E = 1.69^{+0.59}_{-0.45}$,][]{Yang2019} and Sgr A$^*$ \citep[$\alpha_E = 1.65\pm0.02$,][]{Wang2015,Li2015}.
Some FRB theoretical models are related to giant pulses of pulsars \citep[e.g.][]{Cordes2016}. The giant pulses of Crab pulsar also show a power-law distribution \citep[$\alpha_E = -1.85\pm0.10$,][]{Popov2007,Lyu2021}. A power-law distribution of energy is a natural predication of self-organized criticality theory \citep{Bak1987,Aschwanden2011}.

The energy distribution in \gqq{the} low-energy range deviates from this power-law. This deviation was also reported by \citet{Gourdji2019}.
They attributed this deviation to the incompleteness at low energies. In our sample, the turning point is about $10^{38}$ erg, much larger than the threshold energy, which is about $3\times 10^{37}$ erg \citep{Li2021}. 
Therefore, this deviation is physical. 
A similar deviation was also found in other samples \citep{Wang2019ApJ...882..108W,Oostrum2020AA...635A..61O,Cruces2021MNRAS.500..448C}. 
\citet{Li2021} also investigated the energy distribution of FRB 121102. They found the bimodal energy distribution, \gq{which is the Cauchy function with $\alpha_E = 1.85\pm0.3$
for $E > 10^{38}$ erg plus the log-normal function with $\sigma_E = 0.52, N = 2.06\times10^{38}, E_0 = 7.2\times 10^{37}$ erg for $E < 10^{38}$ erg.}
They also give the KDE of energy and burst time in the panel (c) of their Figure 1. In this panel, the high-energy component is only visible in
early observations, while the low-energy peaks are present in both early and late observations.
This novel structure may suggest different physical mechanisms for high-energy and low-energy bursts.
If this conjecture is true, it can explain the deviation in low energies. \gq{The distribution also shows an obvious steepening above $E>5 \times 10^{39}$ erg, which indicates the burst maximum energy $E_{\rm max}$ is around this point. }

\subsection{Waiting time Distribution} \label{subsec:weibull}
\begin{figure}
    \centering
    \includegraphics[width = 0.8\linewidth]{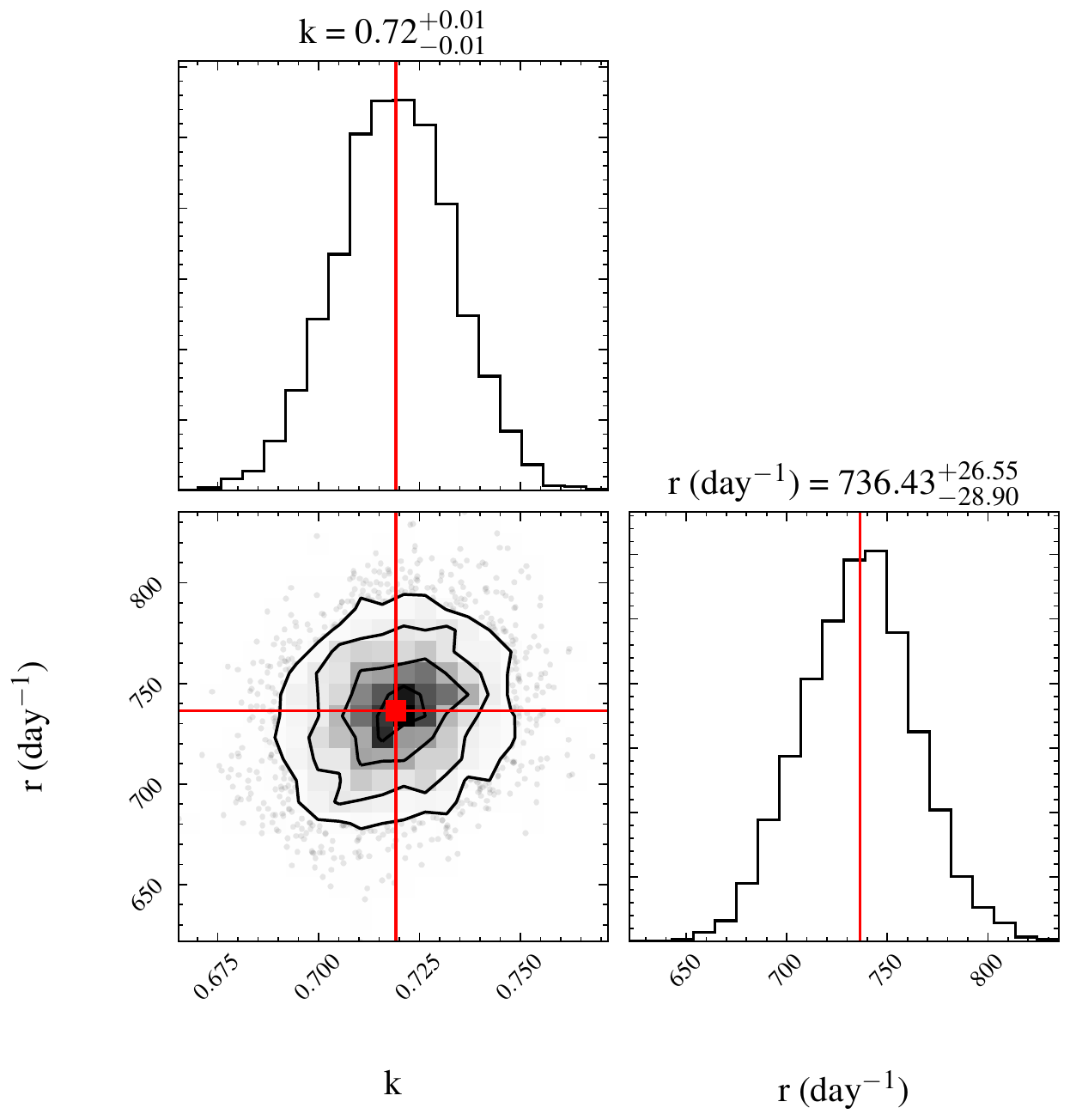}
    \caption{The corner plot for the shape parameter $k$ and event rate $r$ for the Weibull distribution. The waiting times with $\delta_t < 30$ ms and $ \delta_t > 0.5 $ day are excluded. \gq{We use the red lines to denote the best-fitting results, which} are $ k = 0.72^{+0.01}_{-0.01} $ and $r = 736.43^{+26.55}_{-28.90}$ per day.}
    \label{fig:weibullfit}
\end{figure}

\begin{figure}
	\centering
	\includegraphics[width = 0.8\linewidth]{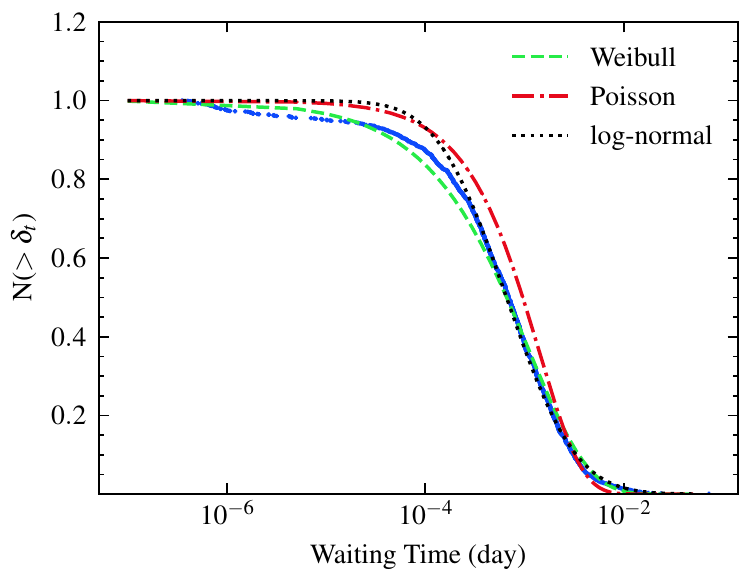}
	\caption{\gq{The cumulative distribution of the waiting time between consecutive bursts}. The blue scatter is the burst waiting time distribution. The dashed green
	line, dot-dashed red line, and dotted black line are best-fitting results of Weibull distribution, Poisson distribution, and log-normal distribution, respectively. \gq{The best-fitting results are  $ k = 0.72^{+0.01}_{-0.01} $ and $r = 736.43^{+26.55}_{-28.90}$ per day for Weibull distribution, $r = 722.90$ per day for Poisson distribution, and $\mu = -7.33, \sigma = 1.26$ for log-normal distribution. The $\chi^2$ for Weibull function, Poisson function and log-normal function are $346.50, 392.28$ and $394.90$, respectively.}}
	\label{fig:wcdf}
\end{figure}

\gq{\citet{Wang2017} found that the time interval between consecutive bursts does not follow a Poission distribution. Afterwards, \citet{Oppermann2018MNRAS.475.5109O} found \gqq{that} the Weibull function is a better description of waiting time,}
which is \gqq{described by}
\begin{equation}
    \label{eq:weibull}
    \mathcal{W}(\delta_t \mid k, r)=k \delta_t^{-1}[\delta_t r \Gamma(1+1 / k)]^{k} \mathrm{e}^{-[\delta_t r \Gamma(1+1 / k)]^{k}},
\end{equation}
where $\delta_t$ is the waiting time, $k$ is the shape parameter, $r$ is the mean burst rate, and $\Gamma$ is the gamma function.
The case $k = 1$ corresponds to the Poisson distribution. Previous works suggested
that the repeating behavior of FRB 121102 tends to have $k < 1$ \citep{Oppermann2018MNRAS.475.5109O,Oostrum2020AA...635A..61O,Cruces2021MNRAS.500..448C},
which means that time clustering is favored. We use the Weibull function
to fit the burst behavior. The observations of FAST lasted for \gqq{roughly} 1 hour per day. Therefore, waiting time greater than a half day
is caused by the observing window, which is ignored in our analysis. The waiting time with $\delta_t < 30$ ms is also excluded.
It is difficult to determine whether they are two separate bursts or multiple peaks of a single burst \citep{Cruces2021MNRAS.500..448C}. We use the MCMC (Markov chain Monte Carlo) technique
to fit this distribution. The corner plot of the best-fitting results is shown in Figure \ref{fig:weibullfit}.
We list the best-fitting parameters in Table \ref{tab:weibull}, \gq{alongside the results from previous works}.
We find $k = 0.72^{+0.01}_{-0.01}$ (1$\sigma$ uncertainty) and $r = 736.43^{+26.55}_{-28.90}$ day$^{-1}$ (1$\sigma$ uncertainty). This burst rate is much higher than the rates derived in previous 
works, which is caused by the high sensitivity of FAST. \gqq{The 90\% threshold of FAST sample is $2.5\times 10^{37}$ erg \citep{Li2021}.}. 
The cumulative distribution of waiting time is shown in Figure \ref{fig:wcdf}. The green dashed line is the predicted result of the Weibull function, the dot-dashed red line is predicted by a Poisson process, \gq{and the black dotted line is the best-fitting results of the log-normal distribution.
Some previous works suggest log-normal distribution is suitable to fit the waiting time distribution \citep{Gourdji2019,Katz2019MNRAS.487..491K}. 
\citet{Li2021} used two log-normal functions to fit the waiting time distribution. We add the fit of log-normal distribution as a reference and do not discuss it in detail.
The $\chi^2$ values for Poisson distribution, Weibull distribution, and log-normal distribution are 392.28, 346.50, and 394.90, respectively. The Weibull function is the best in fitting the data. 
In Figure \ref{fig:wcdf}, Poisson and log-normal distributions cannot well describe short waiting times. However, the Weibull distribution has a better description of that.
For long waiting times, both the Weibull distribution  and log-normal distribution can model the data well.}

\begin{figure}
    \centering
    \includegraphics[width = 0.8\linewidth]{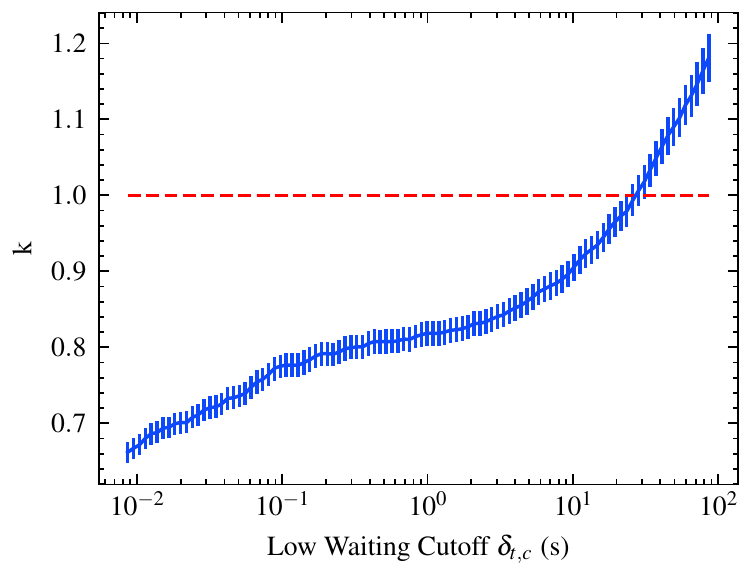}
    \caption{The shape parameter $k$ against the low-waiting-time cutoff $\delta_{t,c}$ for Weibull distribution. The best-fitting results of the
    Weibull function with $ \delta_t > \delta_{t,c} $ are derived. We select some different $ \delta_{t,c} $
    and show the shape parameters $ k $ in blue lines. The blue vertical lines are 1$ \sigma $ uncertainties. The dashed red line is $k = 1$. We found 
    that the burst behavior \gq{approaches} a Poisson process when $ \delta_{t,c} = 28 $ s (\gq{red-dashed line}).}
    \label{fig:kvstc}
\end{figure}

\citet{Oppermann2018MNRAS.475.5109O} collected 17 bursts observed by Arecibo, Effelsberg, GBT, VLA, and Lovell. They derived $k = 0.34^{+0.06}_{-0.05}$ with these 17 bursts.
\citet{Oostrum2020AA...635A..61O} obtained $k = 0.49 \pm 0.05$. Their results are smaller than ours. This may be caused by the low waiting time cutoff.
\citet{Cruces2021MNRAS.500..448C} found that when excluding the waiting time $\delta_t < 1$ s, the shape parameter $k$ shifts from $k = 0.62^{+0.10}_{-0.09}$ to $k = 0.73^{+0.12}_{-0.10}$.
We carefully check the effect of waiting time cutoff $\delta_{t,c}$  and show the dependence of $k$ on $\delta_{t,c}$ in Figure \ref{fig:kvstc}. The solid blue line is the
evolution of $k$ and the vertical solid blue lines are 1$\sigma$ errors. The shape parameter $k$ increases as $\delta_{t,c}$ increases. When $\delta_{t,c} \simeq 28~ s$, the shape parameter $k$ closes to 1, which means that
the repeating behavior is due to a Poisson process. \citet{Cruces2021MNRAS.500..448C} also found that the waiting time with $\delta_t > 100 $ s is consistent with a Poisson process, which is similar to our results. 

\subsection{Linear repetition pattern}
\begin{figure}
    \centering
    \includegraphics[width = 0.8\linewidth]{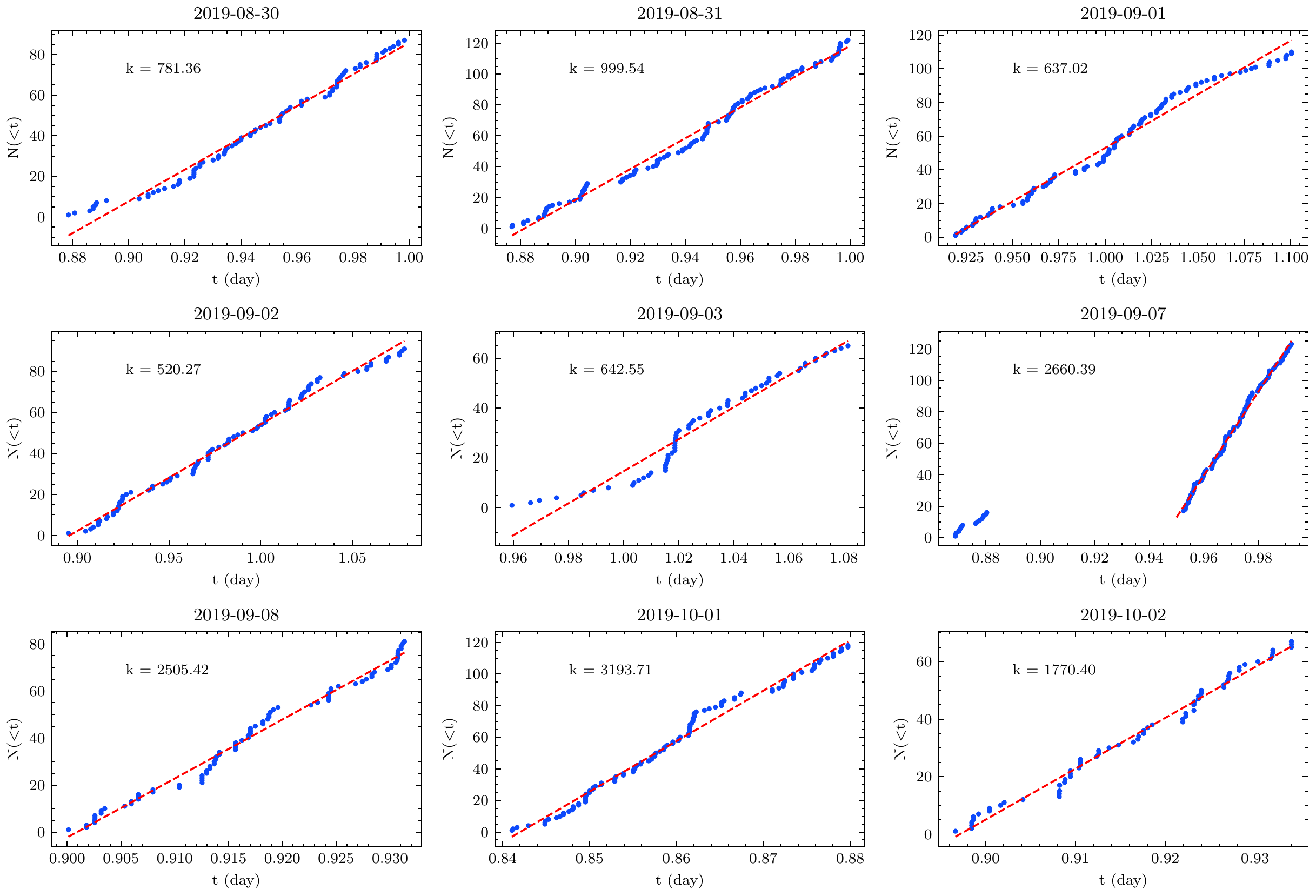}
    \caption{The cumulative distribution of burst number in single days. We select the days with the number of bursts greater than 60. \gq{The x-axises are the decimal part of MJD burst time.}
    	In all cases, the distribution can be well fitted with $N(<t) = st + b$, \gq{where $ t $ is the burst time, and $ s, b $ are fitting parameters. $ N(< t) $ is the number of bursts with burst time earlier than $ t $.} The best-fitting results are shown as dashed 
	red lines in each panel.}
    \label{fig:Nt}
\end{figure}

We use the FAST data to check the burst repetition pattern. We select the single-day observations with the number of detected bursts larger than 60 and show the cumulative distribution function \gq{of the burst time} with blue scatters in each panel of Figure \ref{fig:Nt}. The 
observation time is listed in \gq{the top of} each panel. \gq{The x-axis is the decimal part of the MJD time.}
We find that the 
distribution can be fitted with a linear function
\begin{equation}
	\label{eq:Ntk}
	N(<t) = s t + b,
\end{equation}
\gq{where $t$ is the burst time, $s$ is the slope, and $b$ is a constant.}
The best-fitting results with the dashed red lines are \gqq{shown} in each panel and the best-fitting parameters are also listed. 
\gq{The observation on 2019-09-07 is slightly different from other observations. There is an apparent temporal gap between early bursts occurring before MJD 58732.90
and late ones occurring after MJD 58732.90. In this observation, most bursts occurred after MJD 58732.90 and these bursts can be fitted with Equation (\ref{eq:Ntk}).
For all the observations listed in Figure \ref{fig:Nt}, } although the fitting results are different, they can be all fitted with equation \ref{eq:Ntk}. 
\citet{Tabor2020ApJ...902L..17T} proposed that the burst rate is constant per logarithmic time \gq{with the observation of GBT \citep{Zhang2018ApJ...866..149Z}.}
They found that the cumulative distribution function can be fitted with $N(<t) = \alpha \ln(t/t_0)$. \gqq{Contrary, our results suggest that FAST strongly favors a linear pattern.}
\gq{\citet{Tabor2020ApJ...902L..17T} excluded the model that repeating FRB sources origin from pulsars \citep{Beniamini2020MNRAS.496.3390B}. According to our results, this model is still feasible. The linear pattern may suggest a potentially short period. \gqq{However, similar to \citet{Li2021}, we do not find any short period. The Lomb-Scargle method and epoch folding method are used to search between  10 milliseconds and 
30 minutes. No significant period was found.
The fluctuations of data points along the fitted lines in Figure \ref{fig:Nt} may suggest that the linear 
pattern is caused by the narrow waiting time distribution. Besides, the slopes in varied observations are different, which may suggest diverse waiting time distribution in different observations.}}

\section{Discussion}
\subsection{Energy distribution in different epochs} \label{subsec:enedis2}
\begin{figure}
    \centering
    \includegraphics[width = 0.8\linewidth]{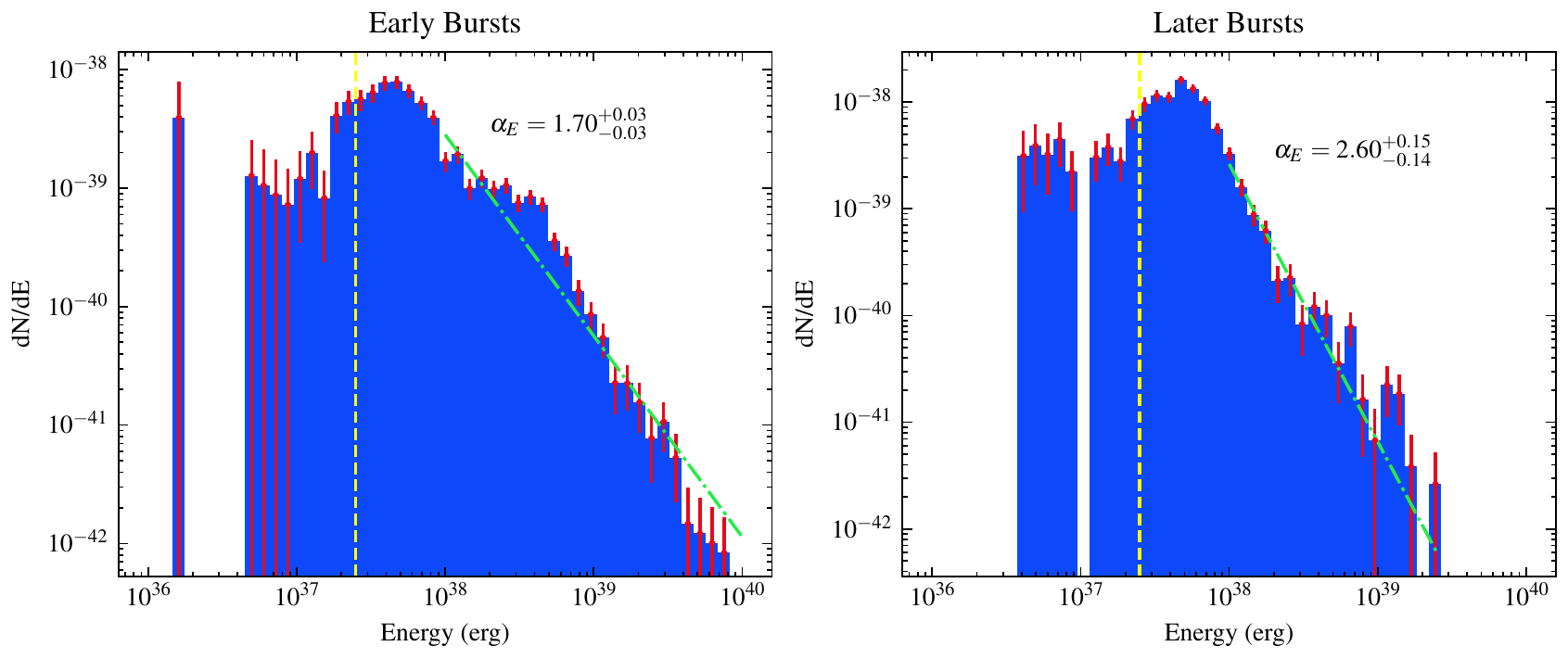}
    \caption{The energy distribution of bursts in different epochs. \gq{The early and late bursts correspond to the bursts occurring before and after MJD 58740, respectively.}
    \gq{The vertical yellow dashed is the 90\% threshold of FAST telescope.} We use the simple power law to fit the bursts with energy greater than $10^{38}$ erg.
    The dot-dashed green lines are the best-fitting results. The power-law indices are also shown in each panel.}
    \label{fig:twopowerlaw}
\end{figure}

An important conclusion of \citet{Li2021} is the novel triple peaks structure in the KDE of energy and burst time (As shown in the panel (c) of their Figure 1). There is a clear energy gap
between the high-energy bursts and low-energy bursts \gqq{before MJD 58740}. The temporal gap between early bursts and late bursts is also significant. Following the 
division of \citet{Li2021}, we regard the bursts occurring before MJD 58740 as early bursts and the bursts occurring after MJD 58740 as late bursts.
The energy distributions \gq{of early bursts and late bursts} are shown in Figure \ref{fig:twopowerlaw}. Again, we only fit the bursts with energy $> 10^{38}$ erg. The best-fitting results 
are shown as green dot-dashed lines. The power-law index for the early bursts is $-1.70^{+0.03}_{-0.03}$ (1$\sigma$ uncertainty), while that for the late bursts is $-2.60^{+0.15}_{-0.14}$ (1$\sigma$ uncertainty), which is \gqq{significantly} different. 

Some previous works support the power-law index $ -1.7$ \citep{Wang2017,Law2017,Wang2019ApJ...882..108W}, which is consistent with the 
result of early bursts. However, there are also some works \gqq{suggest} a steeper index \citep{Gourdji2019}, which is close to the result of late bursts. 
This difference may be caused by the following reasons. There may be different modes of burst emission operating in different epochs, so that the observed energy distribution may depend on the observing time. The two different modes may be related to different mechanisms or different emission sites. \citet{Li2021}
reported the triple peaks in the distribution of energy and burst time (as shown in panel (c) of their Figure 1). At the early epoch, the bursts with $E > 10^{38}$ erg belong to the high-energy burst component. However, the high-energy peak is invisible in the late epoch. The bursts with $E > 10^{38}$ erg in the late epoch are an extension of \gqq{the low-energy peak}. If the high-energy bursts and low-energy 
bursts originate from different mechanisms or different sites, which operate in different epochs, the variance of the energy distribution is naturally interpreted.

\subsection{Waiting time for different energies}\label{subsec:waiting3Class}
\begin{figure}
    \centering
    \includegraphics[width = 0.8\linewidth]{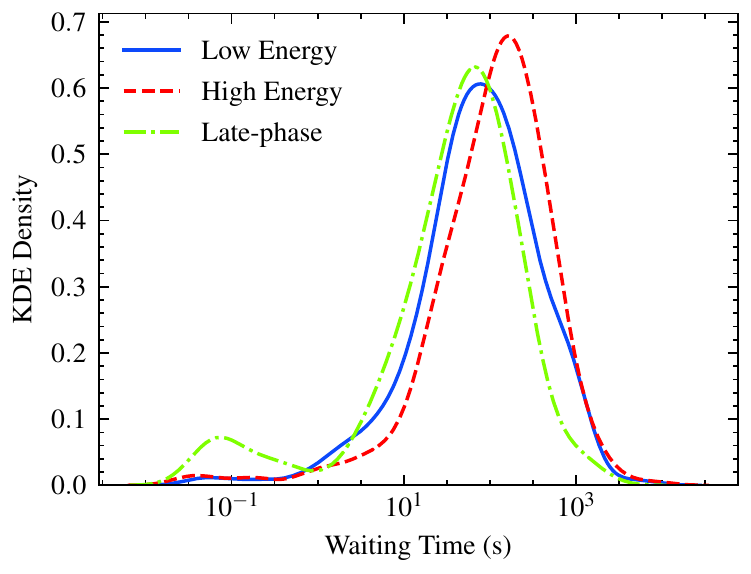}
    \caption{ The KDE of waiting times for three sub-samples. According to the bimodal structure in \citet{Li2021}, we divide 
    the bursts into 3 sub-samples: the low-energy bursts and high-energy bursts from \gq{MJDs 58717 to 58740}, and the late-phase bursts. \gq{The distributions of three sub-samples 
are similar, but with different median values, which are 78.19 s, 133.61 s and 54.62 s for low-energy bursts, high-energy bursts and late-phase bursts, respectively.}}
    \label{fig:3PDFWaiting}
\end{figure}

\citet{Li2021} reported the triple peak structure in the panel (c) of their Figure 1. 
The bimodal structure in the energy distribution in the early phase (\gq{before MJD 58740}) may suggest that they are produced by different physical processes. Therefore, we \gq{explore} whether there
are differences in their other properties, \gq{such as waiting time}. If the low-energy bursts and high-energy bursts are produced by different processes, the waiting time should be
calculated independently. We divide all bursts into three sub-samples: the low-energy bursts and high-energy bursts between MJD 58717 and 58740, and the late-phase bursts detected between MJD 58740 and 58776. The dividing line of burst time is consistent with the choice of \citet{Li2021}.
We take the dividing line between the low-energy bursts and high-energy bursts as $1.58\times10^{38}$ erg.
\gq{The gap is about $10^{38.1} \sim 10^{38.3}$ erg. \gqq{For convenience, we take the separation line as $10^{38.2} \simeq 1.58\times 10^{38}$ erg.}
We also test other choices of the criteria and \gqq{find that it does not affect on our result.}}
For each sub-sample we calculate the waiting time independently and show the \gq{KDE of the waiting time} in Figure \ref{fig:3PDFWaiting}.
\gq{The KDE for these three sub-samples is \gqq{slightly} different. The medians of the low-energy bursts, high-energy bursts, and late-bursts are 
78.19 s, 133.61 s, and 54.62 s, respectively.
There is a small peak in the KDE density of the late-phase bursts, which is close to 0.1 s. }

\begin{figure}
	\centering
	\includegraphics[width = 0.8\linewidth]{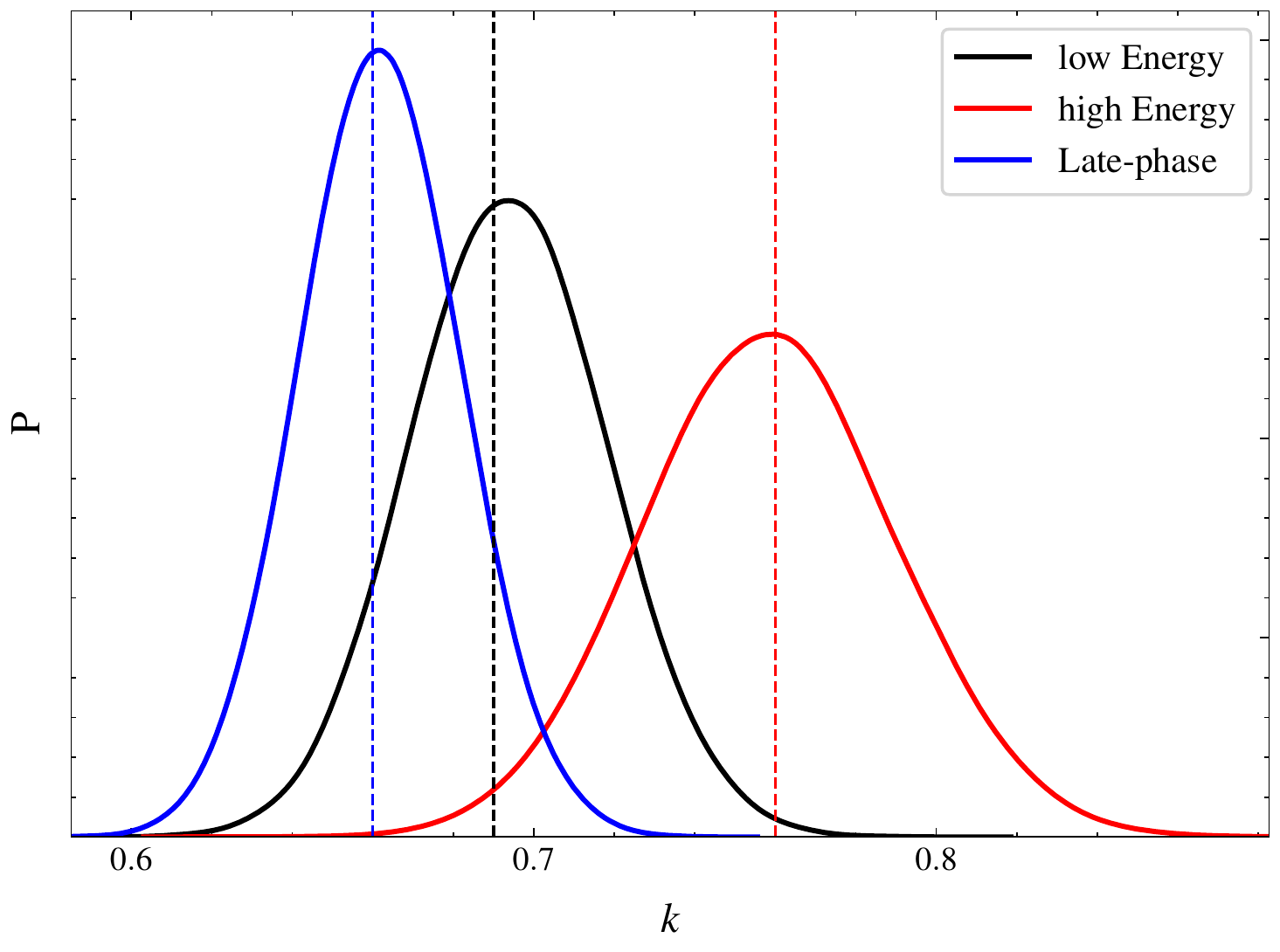}
	\caption{The shape parameter $k$ for the three sub-samples. The solid lines are the posterior probability distributions of the $k$ parameter for the three sub-samples, and the dashed lines indicate the mean values of the posterior distribution. The value of $k$ for low-energy bursts and late-phase bursts are consistent. However, the value of $k$ for high-energy bursts is slightly higher.}
	\label{fig:3k}
\end{figure}

We also use the Weibull function to fit three sub-samples. Again, we ignore the waiting times with $\delta_t < 30$ ms and $\delta_t > 0.5$ day. We list the best-fitting results for three sub-samples in Table \ref{tab:weibull}.
The shape parameters are $k \simeq 0.69 \pm 0.02$ for the low-energy bursts, $k \simeq 0.76 \pm 0.03$ for the high-energy bursts and $k \simeq 0.66 \pm 0.02$ for the late-phase bursts. The posterior distributions of $k$ for the three sub-samples are shown in Figure \ref{fig:3k}.
The $k$ parameters for the low-energy bursts, high-energy bursts, and late-phase bursts are shown as black line, red line, and blue line, respectively.
The vertical dashed lines are the best-fitting results.
The shape parameters $k$ for the low-energy and late-phase bursts are consistent with each other within 1$\sigma$ error, while the $k$ for the
high-energy bursts is higher. 
The consistency of the low-energy bursts and late-phase bursts may suggest that they have the same origin, while the high-energy bursts may have a different origin. The difference between the late-phase
bursts and high-energy bursts is significant, but the variance between the low-energy bursts and 
high-energy bursts is insignificant. This may be caused by the rough $1.58\times10^{38}$ erg division line of low-energy bursts and high-energy bursts. 
The count
distributions of low-energy and high-energy bursts are close to a log-normal distribution. They can be extended to include each other. Therefore, some FRBs may be misclassified. If we can make a robust classification, the difference in waiting time may be larger. 

We derive the burst rates for the low-energy bursts and high-energy bursts are $445.35^{+31.85}_{-30.02} $per day and $354.74^{+27.14}_{-26.14}$ per day, respectively. The burst rate of 
the low-energy bursts is higher than that of the high energy bursts. These two classes of bursts occurred in the same epoch. For the bursts occurring in late observation days, the burst rate is about $800.54^{+48.84}_{-48.32}$ per day, which is higher than that of low-energy bursts 
and high-energy bursts, but close to the sum of these two classes of bursts.

\begin{figure}
	\centering
	\includegraphics[width = 0.8\linewidth]{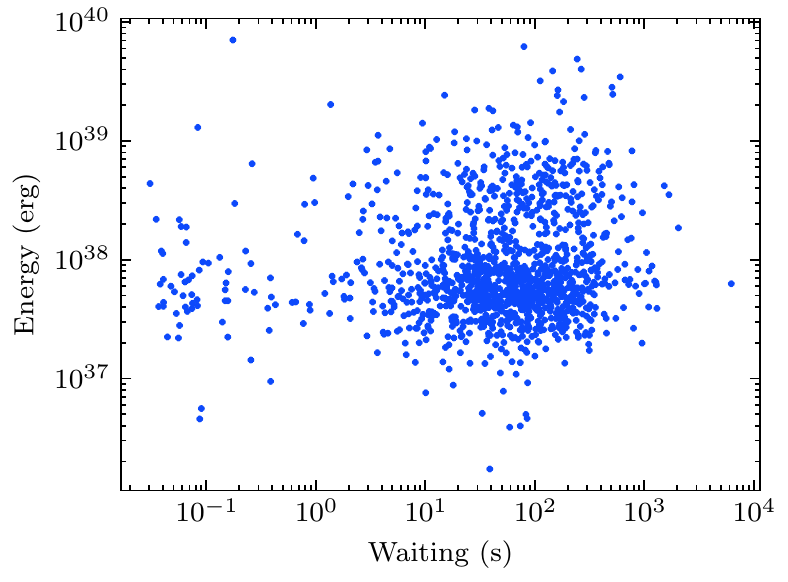}
	\caption{Scatter plot between waiting time and the energy of the burst after the waiting time. 
	There is no correlation between burst energy and waiting time.}
	\label{fig:EvsWaiting}
\end{figure}

We also check the dependence of energy on waiting time. For X-ray flares of gamma-ray bursts, an anti-correlation is found between energy and waiting time \citep{Yi2016}. The scatter plot of energy and waiting time of the FAST FRB sample is shown in Figure \ref{fig:EvsWaiting}.
No significant dependence of energy on waiting time is found. We also check the dependence for three sub-samples and found no significant dependence.

\subsection{Waiting time in different observational days}
\begin{figure}
	\centering
	\includegraphics[width = 0.8\linewidth]{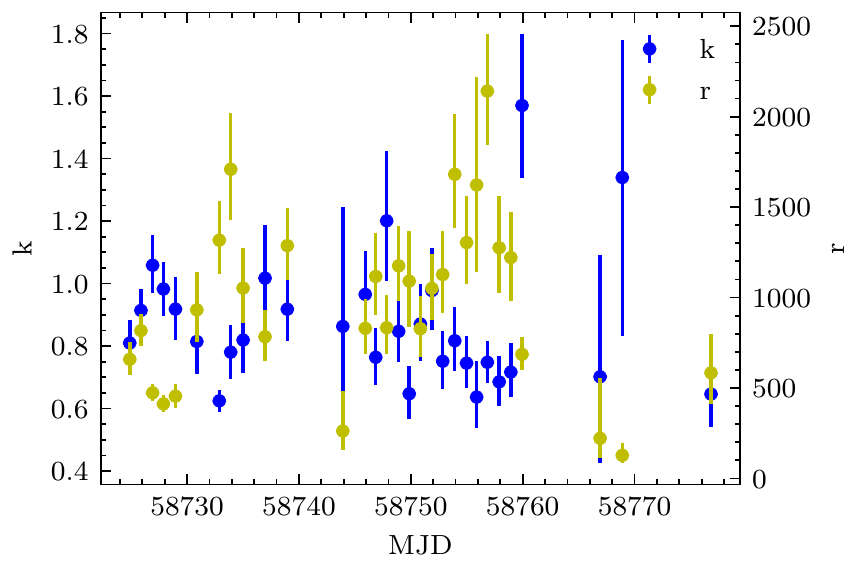}
	\caption{Shape parameters $k$ and burst rate $r$ in single days with the burst number greater than 5. The blue points are the shape parameters $k$ with $1\sigma$ errors, while the yellow points are the burst rate $r$ with $1\sigma$ errors. }
	\label{fig:krevolve}
\end{figure}

The FAST observations spanned two months in time, which covered most of the active window of FRB 121102. We can test the dependence of the
burst behaviors on observing time. The observations lasted for \gqq{roughly} 1 hour each day.
We regard the observations in each day as single samples and select some samples with the number of observed bursts greater than 5.
Then, the Weibull function is applied to fit the waiting time distribution in these days.
The best-fitting results against burst time are shown in Figure
\ref{fig:krevolve}. The blue points are the shape parameters $k$ with $1\sigma$ errors, and the yellow points are the burst rates $r$ with $1\sigma$ errors. 
Interestingly, the value of $k$ is close to $1$ in some days,
which indicates a rough Poisson distribution of the waiting time. \gq{Some observations have $k > 1$, which means that the waiting time prefers a specific value.
If $k\gg 1$, it means a periodic behavior. In our calculation, the value of $k$ is not enough to support the periodic behavior. It only suggests that the waiting time has
a narrow distribution in some days, which is consistent with the linear pattern.} \gqq{The best-fitting result varies for different samples. We do not find significant 
dependence on the observation time.}

\section{Conclusions}
The energy and waiting time distributions of FRB 121102 provide a clue \gq{for understanding} the physical origin of repeating FRBs. 
\gq{\citet{Li2021} reported a large sample of bursts (1652), about 1200 of which are faint bursts with $E < 1.58\times10^{38}$ erg (the nominal threshold for separating low and high energy events).} The detection
of these faint bursts enables us to perform extensive statistical studies of the burst properties. Our conclusions can be summarized as follows.

\begin{enumerate}
	\item The energy distribution in the high-energy range can be fitted with a simple power-law function. The power-law index is $ -1.86\pm0.02 $, which is 
	consistent with previous results \citep{Wang2017,Law2017,Wang2019ApJ...882..108W,Cruces2021MNRAS.500..448C}. 
	\gq{This index is also similar to those of non-repeating FRBs \citep{Cao2018,Lu2019} and magnetar X-ray bursts \citep{Gogus2000,Cheng2020}, which may suggest they share similar physical processes.}
	However, the energy distribution deviates from this power-law function at low energies.
	The turning point is \gq{near} $ 10^{38} $ erg, \gq{well above the 90\% detection completeness threshold ($E_{90} = 2.5\times10^{37}$ erg)}. Thus, this deviation is not caused by the incompleteness in the low-energy end. A different physical mechanism or different emission site for low-energy events may be required to explain this deviation.
	\item The Weibull function is used to fit the waiting times excluding \gqq{events separated by} $\delta_t < 30$ ms and $\delta_t > 0.5$ day. We derive 
	$k = 0.72^{+0.01}_{-0.01}$ and $r = 736.43^{+26.55}_{-28.90}$ day$^{-1}$. This result ($k < 1$) suggests that the bursts of FRB 121102 tend to cluster in time. We also found that when we exclude the waiting time $\delta_t < 28$ s, the burst behavior can be roughly described by a Poisson process.
	\item For bursts observed in a single day, a linear repetition pattern \gq{of burst time} is found. \gq{We select the observational days when more than 60 bursts were detected.}
	The cumulative distributions of the burst time are shown in Figure \ref{fig:Nt}. These distributions can be fitted with $ N(<t) = st + b $. 
	\item The distributions of high-energy early bursts and late bursts are different.\gq{The dividing line of early bursts and late bursts is MJD 58740.} When the bursts with $E > 10^{38}$ erg are considered, the
	power-law index is $-1.70$ for early bursts and $-2.60$ for late bursts. 
	This difference may suggest that they have different physical origins. 
	\item According to the bimodal structure of energy distribution \gq{reported by \citet{Li2021}}, we divide the bursts into 3 classes: the low-energy bursts and high-energy bursts between 
	MJD 58717 and 58740, and the late-phase bursts. \gq{The distributions of waiting time for these three classes are somewhat different. We also use the Weibull function to fit the waiting time distributions.} The posterior probability distribution of the $k$ parameter for the high-energy bursts is slightly different from that of the low-energy and late-phase bursts. These differences may suggest that they have different physical origins or emission sites.
	\item There is no significant dependence of energy on waiting time. For 
	three sub-samples, we also do not find any dependence between the two parameters.
\end{enumerate}

\section*{Acknowledgements}
We would like to thank the anonymous referee for helpful comments. This work was supported by the National Natural Science
Foundation of China (grant No.~11988101, No.~U1831207, No.~11833003,  No.~11725313, No.~11690024, No.~12041303 and No.~U2031117), the Fundamental Research Funds for the Central Universities (No. 0201-14380045), the National Key Research
and Development Program of China (grant~2017YFA0402600); and the National SKA Program of China No. 2020SKA0120200, the Cultivation Project for FAST Scientific Payoff and Research Achievement of CAMS-CAS. PW acknowledges support by the Youth Innovation Promotion Association CAS (id.~2021055) and CAS Project for Young Scientists in Basic Reasearch (grant~YSBR-006). This work made use of data from FAST, a Chinese national mega-science facility built and operated by the National Astronomical Observatories, Chinese Academy of Sciences.

\bibliography{main}{}
\bibliographystyle{aasjournal}

\begin{table}[]
	\centering
	\begin{tabular}{|l|l|l|}
		\hline
		Data set                             & $r$ (day$^{-1}$)             & $k$                      \\
		\hline
		All bursts of \citet{Li2021}                          & $736.43^{+26.55}_{-28.90}$ & $0.72^{+0.01}_{-0.01}$ \\
		\hline
		low-energy bursts                    & $445.35_{-31.59}^{+30.42}$ & $0.69_{-0.02}^{+0.02}$ \\
		\hline
		high-energy bursts                   & $354.74_{-26.14}^{+27.14}$ & $0.76_{-0.03}^{+0.03}$ \\
		\hline
		late-phase bursts                    & $800.54_{-48.32}^{+48.84}$ & $0.66_{-0.02}^{+0.02}$ \\
		\hline
		\citet{Oppermann2018MNRAS.475.5109O} & $5.7^{+3.0}_{-2.0}$        & $0.34^{+0.06}_{-0.05}$ \\
		\hline
		\citet{Oostrum2020AA...635A..61O}    & $6.9^{+1.9}_{-1.5}$        & $0.49^{+0.05}_{-0.05}$ \\
		\hline
		\citet{Cruces2021MNRAS.500..448C}    & $74^{+31}_{-22}$           & $0.62^{+0.10}_{-0.09}$ \\
		\hline
	\end{tabular}
	\caption{The best-fitting results of Weibull distribution for different samples. The classification of the low-energy bursts, high-energy bursts, and 
    late-phase bursts is given in Section \ref{subsec:waiting3Class}. \gq{We exclude waiting times with
    	$\delta_t > 0.5$ day and $\delta_t < 30$ ms from all bursts, low-energy bursts, high-energy bursts, and late-phase bursts. \gqq{The results 
    from other works are listed for reference.}}}
	\label{tab:weibull}
\end{table}

\end{document}